\documentclass[12pt,A4]{article}
\usepackage{epsfig}
\usepackage{array}
\usepackage{amsmath}
\usepackage{amssymb}

\newcommand{\m}{\mathbf}

\newcommand{\be}{\begin{equation}}
\newcommand{\ee}{\end{equation}}
\newcommand{\bea}{\begin{eqnarray}}
\newcommand{\eea}{\end{eqnarray}}
\newcommand{\oh}{\frac{1}{2}}

\begin{document}

\title{Anomalous internal conversion --- the key to understanding the $^{209}$Bi riddle }
\author{ F. F. Karpeshin \\ Mendeleev All-Russian Research
Institute of Metrology, \\ Saint-Petersburg, Russia \\
\bigskip \\
and \\ M.B.Trzhaskovskaya \\
Petersburg Nuclear Physics Institute, Kurchatov Research Center, \\ Gatchina,
Russia }
\maketitle

\abstract{ Theory of anomalous internal conversion is developed, and extended for the description of the hyperfine splitting. Experimental data on the hyperfine splitting in the H- and Li-like heavy ions of $^{209}$Bi are analyzed in terms of the Bohr---Weisskopf effect. Agreement with the theory is achieved, scheduling light on the structure of the magnetization distribution over the nuclear volume. }

\vspace*{1cm}

\newpage

\section{Introduction}

A considerable progress during the past decade was archived in investigation of few-electron heavy ions. Specifically, this concerns study of their electronic structure and its influence on the nuclear processes. First of all, this is electromagnetic decay of the nuclei. Nuclear decay can be enhanced considerably by making use of  the resonance with the electronic transitions, via irradiating with  resonance field of a laser \cite{kabzon,chin1,chin2,book}. Wonderful experiments were performed studying the shell effects on the beta decay \cite{beta}.  There is the comparative study of $\alpha$ decay in H-, He-like ions on the urgent agenda, with respect to that in neutral atoms \cite{alf1,alf2,alf3}. In spite of  that the influence of the electron screening on  $\alpha$ decay is a very important question, in view of many applications in astrophysics and experiments with laser-produced plasma, it is only recently that the adequate approach has been found \cite{alf1}. It was clearly shown that the frozen-shell approximation, which was used during half century, exaggerates the effect of the shell at least  by an order of magnitude \cite{alf2}. Moreover, it gives the wrong sign of the effect \cite{alf3}. Calculations of the $\alpha$ decay rate in solids and powerful magic fields performed within this approach  provide with the basis   for the  wide experimental research \cite{dzub,chin}.

Study of the hyperfine structure in H-like and few-electron heavy ions is of great interest because it makes a stringent test of electronic functions. Moreover, it can provide a test of QED. But the latter must be conducted with proper account of classical theory of internal conversion involving anomalous conversion. The reason is the Bohr-Weisskopf effect \cite{BW}, which comprises a contrinbution of approximately 2\% to the hyperfine splitting. This effect is due to manifestation of the finite distribution of the nuclear magnetic currents within the nuclear volume. Its actual contribution   depends on the nuclear model \cite{HFS}. We remind that the Bohr---Weisskopf effect comprises a contribution to the lowest $1s$ and $2s$ levels in $^{209}$Bi of 2\% and 2.2\%, respectively. This confines that precision with which the nuclear properties can be calculated or predicted, in spite of that the electronic wavefunctions are known with seemingly fantastic accuracy \cite{shab2}.  In order to avoid this difficulty, a method of specific differences was proposed \cite{AD,shab1,barza}.  It was suggested that by making use of the specific differences one can diminish it in the specialy constructed linear combination of the HFS of the two atomic levels. And the parameter $\zeta$ in the linear combination was listed up to five digits \cite{shab2,Loch,Lochp}.
However, the nuclear effects cannot be ruled out  completely \cite{HFS}. Precision of the modern experiment \cite{Loch} already reached such level when discrepancy  with description in terms of  the specific differences became explicitly seen. This discrepancy is  noted in Refs. \cite{Loch, Lochp}, with no explanation. In fact,  such a failure of the specific difference method was  predicted in Ref. \cite{HFS}. The alternative,  nuclear-model independent way of dealing with the Bohr-Weisskopf effect, was offered in \cite{HFS}.  It is based on the consecutive development of the internal conversion theory,  specifically, of bound internal conversion \cite{kabzon,chin1,chin2,book,atta}.
{\it Apriori}, precision of the proposed in {HFS} method of the anomalous moments should be higher than that of the specific differences, as the model-independent method deals with the actual nuclear moments extracted from the same experiment, whereas the specific differences method recipe is founded on a simple conventional Weisskopf model, see below.

 In the following sections, we use the method of the anomalous moments, in order to

1)  Better realize the reasons and mechanism of the ``cancellation''  of the Bohr---Weisskopf effect in the specific differences.

2) To make clear the limits on the precision of the specific differences method.\\
We stress, in order to avoid misleading,  that in no case we subject the results \cite{shab2,Loch,Lochp}, concerning $\zeta$ and $\Delta'E$ values, to any doubts, as they are obtained with electronic functions, which are probably the best at the time being. But we remind at the same time that these values were obtained within a simple nuclear model, and explicitly show the admissible corridor these parameters can take values in, if realistic nuclear models are considered. We show that the experimental data really fall within this corridor. We draw a conclusion about the physical nature of the former discrepancy in the $\Delta'E$ values, which existed between \cite{Loch,Lochp} and \cite{shab2}, and thus solve seeming riddle of experiment \cite{Loch,Lochp}.

\section{Formulae}  \label{formul}

    The general expression for hyperfine splitting, allowing for the Bohr-Weisskopf effect, was obtained in Ref. \cite{HFS} as a series expansion of the diagonal internal conversion matrix element in the limit of the transition energy $\omega \to 0$. This expression shows explicitly that the Bohr---Weisskopf effect is determined by the even $i$-th moments of the magnetization distribution over the nuclear volume, which can consequently be unambiguously retrieved from experiment.  This expression  was obtained using the surface current  (SC) nuclear model \cite{sliv}. Though the expression contains the only model parameter --- radius of the nuclear current  $R_0$, resulting nuclear moments  obtained by means of the analysis of the data are independent of the concrete value of $R_0$. They  hold in numerical variation of the parameter. This model independency was demonstrated in Ref. \cite{HFS} on the example of analysis of data\cite{Loch}.

In order to separate out the Bohr-Weisskopf effect, one can  set the model radius $R_0=0$. This corresponds the no-penetration model in internal conversion theory (\cite{HFS} and refs. therein). The remaining terms contain information about $R_i$.  After the limiting transformation is performed, the resulting expression for the hyperfine splitting (HFS) of the $i$-th level reads as follows:
\be
 w_i=\int_0^\infty g_i(r)f_i(r)dr-\frac3{10} c_1^{(i)}{R_2}^2-\frac3{28}c_3^{(i)}{R_4}^4 +\ldots\,. \label{BW}
\ee
Here $g_i(r)f_i(r)$ are the large an small components of the Dirac electronic wavefunction of the $i$-th level, $i=1$ for the $1s$ level, and $i=2$ for the $2s$ level.   $c_i$ --- the coefficients of the expansion of the sub-integral product of $g_i(r)f_i(r)$ wave functions near the origin \cite{BW,tay,RAINE}.   We remind that  in the nuclear vicinity, the electronic wavefunction is well reproduced by the Taylor series:
\bea g(r) =
a_0+a_2r^2+\ldots, \qquad f(r) = b_1r+b_3r^3+\ldots   \qquad
\text{for the $s$ states}\,,
\label{tay1}    \\
g(r) = a_1r+a_3r^3+\ldots, \qquad f(r) = b_0+b_2r^2+\ldots \qquad
\text{for the $p_\oh$ states}\,.    \label{tay2}     \eea
By making use of (\ref{tay1}), (\ref{tay2}), the
electronic current in (\ref{BW}) can be put down as follows:
\be
j(r) =g_i(r)f_i(r)= c_1^{(i)}r+c_3^{(i)}r^3+ c_5^{(i)}r^5+\ldots\,. \label{ci}
\ee
Coefficients $c_k^{(i)}$  are obtained by means of solving the corresponding Dirac equation, with the help of the package of
computer codes RAINE \cite{RAINE}.
 Fermi nuclear charge
distribution was supposed, with typical parameters.
A conventional value of $R_0 = 1.2A^{\frac13}$
= 7.121 fm was  adopted for the radius of the nuclear charge distribution, $A$ being the mass number of the nucleus. The electron interaction in  the case of Li-like atoms and QED corrections in \cite{HFS} were not involved in the electronic wave function, but added separately within the frame of the perturbation theory.

      The set  of coupled equations (\ref{BW}) is solved by fit  to experimental data. Baring in mind that data \cite{Loch} were obtained for the two levels, $1s$ and $2s$, we put down explicitly two terms for description of the Bohr---Weiskoipf effect, and neglect the remaining terms in our further consideration, in accordance with what is said in Ref. \cite{HFS}. As a result,  one can find nuclear magnetization moments $R_2$, $R_4$, under condition that the coefficients and experimental data are correct.

    Within the SC model \cite{sliv},   the transition current $J(R)$ is
\be
      J(R)\,=\, D\delta (R)\, \qquad  \text{--- SC nuclear model} \,.
\ee
All the moments $R_i\equiv R_0$.
In another, ``opposite'' nuclear  model of volume transition currents (VC) \cite{HF}, assuming
\be
J(R)\,=\, \left\{   \begin{array} {l@{\hspace{2cm}}l}
      const,   & \qquad R\leq R _0   \\
            &\\
            0\,, & \qquad R\geq R _0
        \end{array}
   \right.
   \, \qquad \text{ --- VC nuclear model}
   \,,
\ee
magnetization moments ${R_2}^2=\frac23 {R_0}^2$, ${R_4}^4=\frac12 {R_0}^4$, etc., generally ${R_i}^i = \frac4{4+i}{R_0}^i$.

In order to make the  underlying physics more transparent,  first,  temporarily consider reduced set of equations (\ref{BW}) up to the terms with the second nuclear moment and the first coefficients in the series expansion $c_1$. Then these last terms only contain information about the Bohr---Weisskopf effect. One can readily get rid of  it in the linear combination, called specific difference:
\be
\Delta' E= w^{2s}-\zeta w^{1s}  \label{sde}
\ee
with an evident solution for the nuclear model independent approximation, $\zeta_0$, to the parameter $\zeta$:
\be
\zeta_0= \frac{c_1^{2s}}{ c_1^{1s}}\,.             \label{c1}
\ee
In this approximation, parameter $\zeta_0$ can be calculated with maximal precision with which the electronic functions are known.

      In fact, this model independence is immediately broken with account of the next, that is
      the second term of the series expansion in (\ref{BW}). Resulting approximation $\zeta_2$ to the parameter $\zeta$ becomes
\be
\zeta_2=t_{2s}/t_{1s}, \qquad t_i=\frac3{10}c_1^{(i)} {R_2}^2+\frac3{28}c_3^{(i)} {R_4}^4  \,.
\ee
Estimation of the effect of allowance for the fourth moments can be obtained by substitution of the coefficients found in ref. \cite{HFS}.

    For the present purposes, it is enough to use the coefficients in the approximate wavefunctions, calculated in Ref. \cite{HFS}, in order to access the range of permissible values of $\zeta$ depending on the model used. The coefficients are presented in Table 1. These result in the value of (\ref{c1})  $\zeta_0$ = 0.17549, which is independent on the nuclear model and therefore, the same for the SC and VC models. This value is listed in Table 2. In the next approximation, involving  two terms in the series (\ref{BW}),  with allowance for the $c_2$ coefficients and the $R_4$ moments, the parameter $\zeta_2$ becomes model-dependent. Its values obtained within the SC and VC models are also consecutively listed in Table 2. As one can see from the presented values, account of the second terms diminishes the $\zeta_2$ value  as compared to $\zeta_0$ by approximately 3.2$\times10^{-4}$ in the SC model, and by  2.3$\times10^{-4}$ --- in the VC model. It is important that within the VC model parameter $\zeta$ turns out to be by approximately 9.1$\times10^{-5}$ higher than in the SC model. Fast divergence follows the performed consideration.

\section{Analysis of the experimental data for $^{209}$Bi }

      A little lower  value of $\zeta$ = 0.16886 was reported in \cite{shab1,Loch,Lochp}.  For the specific difference in Ref. \cite{shab2} a value of $\Delta_{\text{theor}}$ = $-$61.320 meV was obtained. Experimental HFS in both $1s$ and $2s$ levels was studied in refs. \cite{Loch,Lochp,exca}. The results disagree with one another beyond the error bars. We will analyze the later data \cite{Loch,Lochp}. For the sake of completeness, they are presented in Table 1. These were compared to the theory \cite{shab2}, within the specific difference method.  Substitution of these values into Eq. (\ref{sde}) yields in the specific difference
$\Delta' E$ = $-$61.37 meV.

      As we saw in the previous section, specifically in Table 2, with allowance for the fourth nuclear moment, at the transition from the SC model to the VC model, the parameter $\zeta$ increases by 9.1$\times10^{-5}$. As the coefficients (\ref{BW}) together with the parameter $\zeta$ vary regularly with change of the wave function, there is every reason to believe that with allowance for the higher order effects of QED, together with the electronic interactions in the case of Li-like state, will not affect this ratio considerably.
In Ref. \cite{shab2}, the results are derived within the conventional Weisskopf nuclear model. In this model, wavefunctions are assumed to be constant within the radius $R_0$. The nuclear current is defined by the matrix element $\langle \bar \psi (\m r)|\m \gamma |  \psi (\m r)\rangle$, where $\m \gamma$ are the the three space Dirac matrices, $\bar \psi (\m r)= \psi^+ (\m r) \gamma_0$. In terms of the non-relativistic wavefunctions, this matrix element leads to  that of the operator $(\m \sigma \m p)$, where $\m p=-i\nabla$ is the momentum operator. Therefore, the Weisskopf model may be considered as of the surface character.  Therefore, by direct  analogy with section \ref{formul}, turning to the VC model, one can anticipate a resulting $\zeta$ value of  2.6$\times10^{-4}$ higher, which will comprise $\zeta_{\text{teor}}^{\text{VC}}$ = 0.16888.   Applying this higher value to the data \cite{Loch,Lochp} in Table 1, we arrive at the specific difference value of $-$61.461 meV, which value can be conditionally considered as the upper bound, as explained above. And the value $\zeta_{\text{W}}$ = 0.16886 therefore becomes like the lower theoretical bound now, also the related specific difference value $\Delta'E_{\text{theor}}$ = $-$61.320 meV is. We see that the discrepancy goes away. Instead, experimental data \cite{Loch,Lochp}, $\Delta_{\text{exp}}$ = $-$61.373 meV is just between the values given by these opposite in their physical sense models. It is closer to the SC model. This is fairly reasonable from the simple consideration on the physical ground. SC model \cite{sliv} has a physical justification in the Pauli principle, which suppresses nucleon motion inside the nucleus, and less suppresses it on the nuclear surface.

\begin{table}
\caption{\footnotesize  Coefficients of then series expansion of the electronic current (\ref{ci}), as calculated in Ref. \cite{HFS}, together with the experimental data concerning HFS  }
\begin{center}
\begin{tabular}{c|c|c|c}
Electronic level & $c_1$, fm$^{-2}$  &  $c_3$, fm$^{-4}$ & $w^{\text{exp}}$, eV \\
\hline
$1s$ & -4.4123$\times10^{-9}$ & 3.7255$\times10^{-11}$ & 5.0863 \\
$2s$ & -7.7431$\times10^{-10}$ & 6.5494$\times10^{-12}$ & 0.7975    \\
\hline
\end{tabular}
\end{center} \label{coeft}
\end{table}

\begin{table}
\caption{\footnotesize Consecutive approximations $\zeta_0$ and $\zeta_2$ to the parameter $\zeta$, calculated with the coefficients from Table 1, and the resulting volume-to-surface nuclear model ratio }
\begin{center}
\begin{tabular}{c|c|c|c}
Approximation to $\zeta$ & $\zeta_{SC}$ & $\zeta_{VC}$ & $\zeta_{VC}/ \zeta_{SC}$     \\
\hline
$\zeta_0$ & 0.175489 & 0.175489 &   1   \\
$\zeta_2$  &  0.175433  & 0.175449  &  1.00010  \\
\hline
\end{tabular}
\end{center} \label{sdet}
\end{table}

\section{Discussion}

We applied results and the nuclear-model independent method of anomalous nuclear moments  developed in \cite{HFS},  deploying it in finer detail, aimed at analysis of experimental data \cite{Loch,Lochp}. The method is based on the set of the coupled equations (\ref{BW}).
From the more general approach provided by this method we showed explicitly the limits within which the specific difference method can be considered as  independent of the nuclear model, ---   up to fourth decimal, which is already lower than the experimental precision. As a result, that model failed to explain the data, in spite of the excellent electronic wavefunctions \cite{shab2}, probably the best presently. On the other hand, experimental data \cite{Loch,Lochp} received a reasonable explanation when the method of anomalous nuclear moments was involved. Use of the parameter $\zeta$ calculated in Ref. \cite{shab2} with high accuracy, allowed us to analyze experiment \cite{Loch,Lochp} on the basis of the method of the anomalous nuclear moments. Actually, this way also offers a stringent test of QED, {\it apriori} more stringent in comparison with the  specific difference method. Moreover, model independence of the way \cite{HFS} may only be  realized if both experimental data are correct, and the electronic functions are precise. This allows one to draw an unusual conclusion that the method can serve as a test for experimental data themselves, not only for theory.  And the main consequence in our opinion is \cite{HFS} that the method can be successfully used for the direct constructive  purpose of retrieving information on the nuclear structure from experimental values of the Bohr-Weisskopf effect, {\it e.g.} as pursued in experiment \cite{perez}, in the spirit of the original paper \cite{BW}, instead of fighting it through the specific difference method.    Such experiments  just appropriate at the contemporary stage of investigations, and can be performed on the storage-ring facilities e.g. in Lanzhou, GSI Darmstadt.   \\

\bigskip
\qquad
    The authors would like to acknowledge many fruitful discussions of the topic with L.F.Vitushkin, V.M.Shabaev, I. I. Tupitsin for  fruitful discussions and helpful comments.


\newpage
\begin{thebibliography}{99}\fussy

\bibitem{kabzon} B.A.Zon,  F.F.Karpeshin,  Zh. Eksp. i  Teor. Fiz., {\bf 97}, 401, 1990 {\it Engl. transl.}     Sov. Phys. --- JETP (USA), {\bf 70}, 224, 1990.


\bibitem{chin1} F.F. Karpeshin, Zhang Jing-Bo and Zhang Wei-Ning,
    Chinese Physics Letters, {\bf 23} (2006) 2391.

\bibitem{chin2} F.F. Karpeshin, M.B.Trzhaskovskaya,
    Zhang Jingbo, Chinese Physics Letters {\bf 23} (2006) 2049; F.F. Karpeshin, M.B. Trzhaskovskaya and J. Zhang, Eur. Phys. J. {\bf A 39}, 341 (2009).

\bibitem{book} F. F. Karpeshin, Prompt Fission in Muonic Atoms and Resonance Conversion. Saint-Petersburg: Nauka, 2006.

\bibitem{beta} M. Jung, F. Bosch, K. Beckert, et al., Phys. Rev. Lett.
{\bf 69}, 2164 (1992);  F. Bosch, T. Faestermann, J. Friese, et al., Phys. Rev. Lett. {\bf 77}, 5190 (1996).

\bibitem{alf1}  F. F. Karpeshin,  Phys. Rev. {\bf C87}, 054319 (2013).

\bibitem{alf2} F. F. Karpeshin, M. B. Trzhaskovskaya, in: Exotic Nuclei, Proc. of the  First African Symposium on Exotic Nuclei, Cape Town, South Africa, 2 -- 6 December 2013. Ed. E. Cherepanov et al., World Scientific: New Jersey---London---Singapore, 2014, p. 201.

\bibitem{alf3}  F. F. Karpeshin, M. B. Trzhaskovskaya, Yad. Fiz. {\bf  78}, 1055 (2015).  ({\it In Russian. Engl transl.:})  Phys. At. Nucl. {\bf  78}, 993 (2015).

\bibitem{dzub} A.Ya. Dzublik, Phys. Rev. C {\bf 90}, 054619 (2014).
\bibitem{chin} Niu Wan, Chang Xu, and Zhongzhou Ren,  Phys. Rev. C {\bf 92}, 024301 (2015).
\bibitem{BW} A.Bohr, V.F.Weisskopf, Phys. Rev. {\bf 77}, 94 (1950).

\bibitem{HFS} F. F. Karpeshin, M. B. Trzhaskovskaya, Nucl. Phys. {\bf A941}, 66 (2015).

\bibitem{shab2} A.V. Volotka, D. A. Glazov, O.V. Andreev, V. M. Shabaev, I. I. Tupitsyn, and G. Plunien, Phys. Rev. Lett. {\bf 108}, 073001 (2012).

\bibitem{AD} J.R. Persson, ADNDT, {\bf 99}, 62 (2013).

\bibitem{shab1} V. M. Shabaev, A. N. Artemyev, V. A. Yerokhin, O. M. Zherebtsov, and G. Soff, Phys. Rev. Lett. {\bf 86}, 3959 (2001).

\bibitem{barza} A. E. Barzakh, L. Kh. Batist, D. V. Fedorov,
    V. S. Ivanov, K. A. Mezilev, P. L. Molkanov, F. V. Moroz, S. Yu.
    Orlov, V. N. Panteleev, and Yu. M. Volkov, Phys. Rev. {\bf C 86},
    014311 (2012).

\bibitem{Loch} M. Lochmann, R. J{\"o}hren, C. Geppert {\it et al.},
arxiv: 1401.8224v1 (2014).

\bibitem{Lochp} Matthias Lochmann,  Raphael J{\"o}hren, Christopher Geppert,   Zoran Andelkovic {\it et al.}, Phys. Rev. {\bf A 90}, 030501(R) (2014).

\bibitem{atta} F.\,F.\ Karpeshin, M. R. Harston, F. Attallah, J. F. Chemin,
     J. N. Scheurer, I. M. Band and M.\,B.\ Trzhaskovskaya, Phys. Rev. C
{\bf 53} (1996) 1640.

\bibitem{sliv} L. A. Sliv, Zh. Eksp. Teor. Fiz. 21 , 770 (1951) (In
Russian).

\bibitem{tay} L.A.Sliv and V.A.Volchok, Preprint FTI Acad. Sci.
    USSR, 1956. (In Russian.)

\bibitem {RAINE}  I. M. Band, M. B. Trzhaskovskaya, C. W. Nestor Jr.,
       P. O. Tikkanen, S. Raman,   Atom. Data and Nucl. Data Tables {\bf 81}, 1 (2002); I. M. Band and  M. B. Trzhaskovskaya,  {\it ibid.} {\bf 55}, 43 (1993); {\bf 35}, 1 (1986).

\bibitem{HF} F.F.Karpeshin and      M.B.Trzhaskovskaya, Hyperfine Interact. 2007,  DOI 10.1007/s10751-006-9506-z; Laser Phys., {\bf  17}, 508 (2007).

\bibitem{exca}P. Beiersdorfer, A. L. Osterheld, J. H.
Scofield, J. R. Crespo Lopez-Urrutia, and K. Widmann, Phys. Rev.
Lett. {\bf 80}, 3022 (1998).

\bibitem{perez} A. Perez, Galvan et al., Phys. Lett. B {\bf 655},  114 (2007).

\end{thebibliography}
\end{document}